\documentclass[11pt,twoside]{article}

\usepackage{asp2006}
\usepackage{epsf}
\usepackage{psfig}
\usepackage{lscape}

\begin{document}
\title{The Physical Properties of LBGs at z$>$5: Outflows and the ``pre-enrichment problem''}
\author{M. D. Lehnert$^1$, M. Bremer$^2$, A. Verma$^3$, L. Douglas$^2$, and N. F\"{o}rster Schreiber$^4$} 
\affil{$^1$GEPI, Observatoire de Paris, $^2$Physics Department, University
of Bristol, $^3$Physics Department, University of Oxford, $^4$MPE}

\begin{abstract}
We discuss the properties of Lyman Break galaxies (LBGs) at z$>$5
as determined from disparate fields covering approximately 500
arcmin$^2$. While the broad characteristics of the LBG population has been
discussed extensively in the literature, such as luminosity functions
and clustering amplitude, we focus on the detailed physical properties
of the sources in this large survey ($\ga$100 with spectroscopic
redshifts). Specifically, we discuss ensemble mass estimates, stellar mass
surface densities, core phase space densities, star-formation intensities,
characteristics of their stellar populations, etc as obtained from
multi-wavelength data (rest-frame UV through optical) for a subsample of
these galaxies. In particular, we focus on evidence that these galaxies
drive vigorous outflows and speculate that this population may solve the
so-called ``pre-enrichment problem''.  The general picture that emerges
from these studies is that these galaxies, observed about 1 Gyr after
the Big Bang, have properties consistent with being the progenitors of
the densest stellar systems in the local Universe -- the centers of old
bulges and early type galaxies.  
\end{abstract}

\section{High Redshift Galaxies and Cosmology}

Galaxies at the highest redshifts, z$>$5, are vital objects in
observational cosmology.  They formed during an era of one of the most
dramatic phase changes in the history of the Universe -- the epoch of
reionization.  Remarkably, these earliest galaxies were able to dictate
the phase of the general medium in which all galaxies are imbedded.
Because of their enhanced [$\alpha$/Fe] ratios, which are the product
of core collapse supernova enrichment, and old ages, the densest
(i.e., globular clusters) and most massive stellar systems (i.e.,
giant ellipticals) are thought to have formed at such high redshifts
in spectacular bursts of star-formation of relatively short duration
($<$100 Myrs to a 1 Gyr).  The short duration of the star-formation
but the large accumulated mass must mean that these galaxies had very
high star-formation intensities  -- the rate of star-formation per unit
surface area.  It is well-known that galaxies with high star-formation
intensities drive vigorous outflows of metals and energy.  Such outflows
from high redshift galaxies could have both cleared material from the
galaxian surroundings allowing ionizing photons to escape as well as
enriching the intergalactic medium (IGM).  Through this mechanism problems
of how the IGM was re-ionized and enriched in metals even at the highest
redshifts yet observed (this is the so-called ``pre-enrichment problem'')
could be solved.  But can we find direct evidence for this scenario in
the physical properties of high redshift galaxies?  This is what we will
attempt to address here.

\begin{figure}[!ht]
\plotfiddle{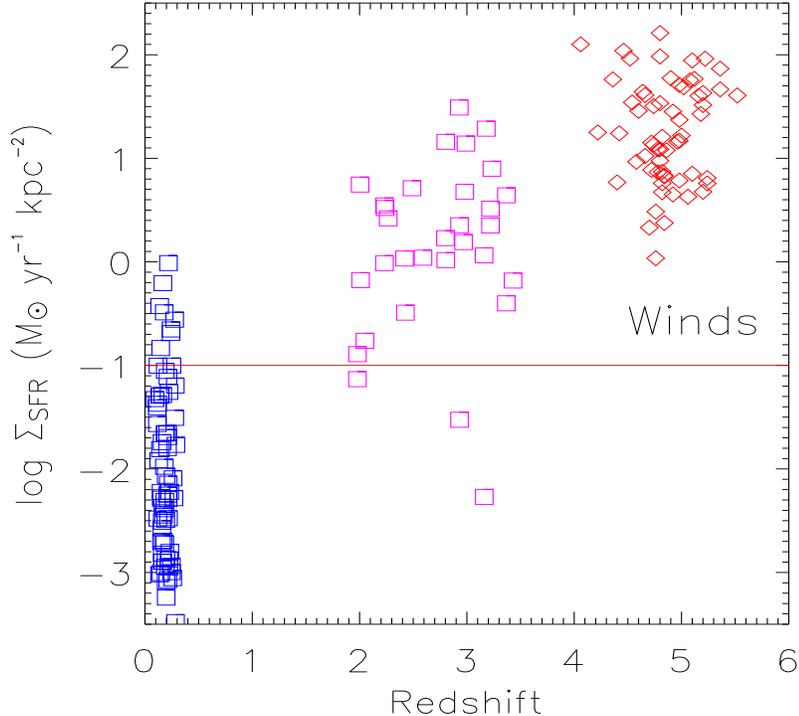}{9.0 cm}{0}{70.0}{60.0}{-230}{-100}
\caption{We compare the star-formation rate intensities -- the rate
of star-formation per unit projected area -- of the z$\approx$5 LBGs (diamonds; Verma et al. 2007), z$\approx$3 LBGs (squares; Papovich
et al. 2001), and a local sample of UV selected galaxies (squares;
Heckman et al. 2005).  The horizontal line indicates the threshold
above which local starburst exhibit large scale outflows (Heckman 2001;
Lehnert \& Heckman 1996).  All of the galaxies studied here lie above
this threshold and thus by analogy, must be driving outflows.  We note
that the evidence is overwhelming that virtually all z$\approx$3 LBGs
are driving outflows (Adelberger et al. 2003).}
\end{figure}

\section{Our Comprehensive Study of z$>$5 Galaxies}

At z$\sim$5, there have been a number of significant spectroscopic surveys
investigating the nature of Lyman-break galaxies (LBGs, or ``V- or R-band
dropouts''), two of which has been lead by some of the authors.
There are now over 100 confirmed redshifts in total at z$\ga$5 through
the work of our ESO Large Program (the ESO Distant Galaxy Survey --
ERGS; Douglas et al. 2007, in preparation) and the study of the ``BDFs''
(e.g., Lehnert \& Bremer 2003; Lehnert et al. 2007, in preparation).
The ESO LP is based on the EDIScS fields, which are 10 widely separated
intermediate-redshift cluster fields, while the ``BDFs'' are four
adjacent fields observable from Paranal in winter. Our total sample
constitutes the largest sample of z$>$5 LBGs (``R-band drop outs'')
with spectroscopically confirmed redshifts.  In all of these fields, we
have deep R-, I-, z-band, and IRAC data (3.6, 4.5, 5.8, and 8.0 $\mu$m).
Some of the fields also have observations in other optical/infrared bands,
including MIPS 24$\mu$m data and HST ACS imaging.  In addition, we have
been studying a similar population of galaxies in the GOODS-S, taking
advantage of the deep multi-wavelength data set (Verma et al. 2007).

\section{What have we concluded?}

From our analysis of some of these extensive data sets (Verma et
al. 2007), we conclude:

$\bullet$ The LBGs at z$\approx$5 are young, with typical ages less than
100 Myrs, compact, with typical half-light radii of about 1 kpc, and
are rapidly forming stars, 10 to 200 M$_{\sun}$ yr$^{-1}$.  These ages
are very young suggesting that they have been intensely forming stars
for only a few dynamical times (estimated from stellar mass estimates
and sizes).  They therefore may likely represent a population of
``primordial galaxies''.

$\bullet$ These sources are generally so young that they likely did
not substantially contribute to reionization.  The estimates of their
ages from SED suggests that they formed at redshifts of about 6-7.
Their contribution to re-ionization would likely be after the Universe
was already substantially ionized.

$\bullet$ From SED fitting from the rest-frame UV through optical,
the LBGs at z$\approx$5 are typically about a factor of 10 less
massive that similarly selected galaxies at z$\approx$3 (few x 10$^9$
M$_{\sun}$ compared to few x 10$^{10}$ M$_{\sun}$; Shapley et al. 2001).
These galaxies represent about 1\% of the local mass density (Verma et
al. 2007).

$\bullet$ The LBGs at z$\approx$5 have high stellar mass surface
densities, $\mu_{\rm stars}$$\approx$2-6 x 10$^8$ M$_{\sun}$ kpc$^{-2}$
and have very roughly approximated core phase densities of 10$^{-6}$
pc$^3$ km$^3$ s$^{-3}$.  Both of these are similar to that of bulges
and spheriods of M$^{\star}$ galaxies at low redshift.

$\bullet$ The LBGs at z$\approx$5 have high star-formation intensities,
well above that needed to drive winds at low redshift, $\Sigma_{\rm
SFR}>>$0.1 M$_{\sun}$ kpc$^{-2}$ yr$^{-1}$ (Figure 1; Heckman 2001;
Lehnert \& Heckman 1996).

$\bullet$ If the z$\approx$5 drive strong outflows then they may be able
to solve the ``pre-enrichment problem'' -- whereby the metallicity of
the IGM does not appear to evolve strongly with redshift at least out
to almost z=6 (Songaila 2001; Ryan-Weber et al. 2006). Quantitatively
(crudely) estimating the likely metal contribution of these galaxies
suggests that they are able to solve the pre-enrichment problem (see
Figure 2).

\acknowledgements 
MDL wishes to thank Johan Knappen for organizing such a delightful meeting
in honor of the contributions of John Beckman and for his extreme patience
in waiting for this contribution.

\begin{figure}[!ht]
\plotfiddle{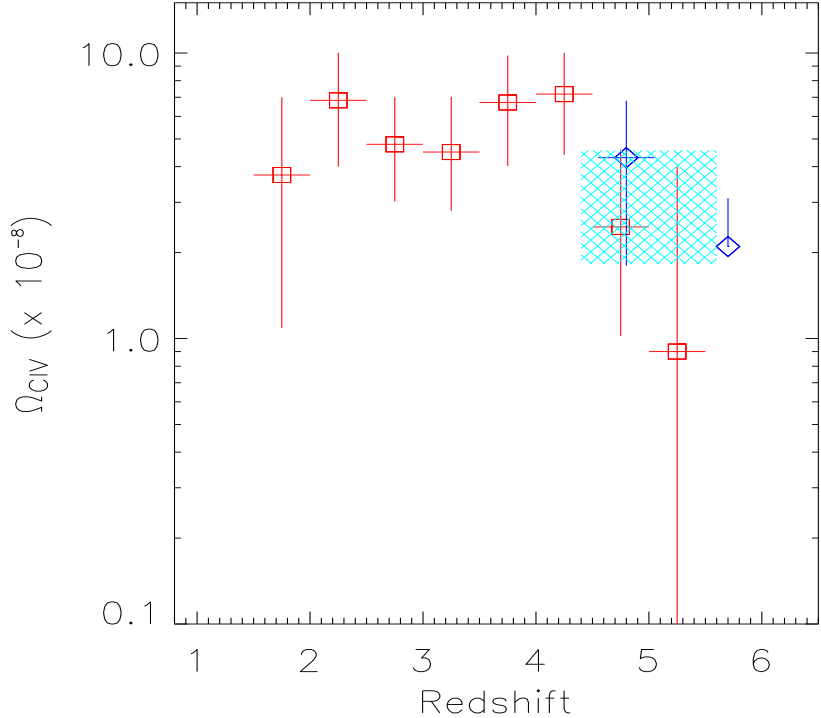}{9.0 cm}{0}{70.0}{60.0}{-220}{-70}
\caption{The co-moving density of CIV absorbers does not appear to
evolve strongly with redshift (Songaila 2001).  This suggests that a
large fraction of the metals in the intergalactic medium may have been
in place within 1 Gyr after the Big Bang.  Comparing the data of the
contribution to the closure density due to CIV absorbers, $\Omega_{CIV}$
(Songaila 2001; squares, and Pettini et al. 2003; Ryan-Weber et al. 2007;
the highest redshift diamond from Ryan-Weber et al. is a lower limit), we
find that it is plausible that the z$\approx$5 LBGs could eject sufficient
metals to explain the IGM absorption line observations.  Our range of
estimates was derived assuming that $\dot{M}_{\rm outflow}$=$\dot{M}_{\rm
star-formation}$ (Lehnert \& Heckman 1996), the closure density, that
the ionization fraction of C$^{3+}$=0.5, the metallicity of the outflow
is 0.2 Z$_{\sun}$, $\Omega_b h^2$=0.023. and the range of star-formation
rates derived in Verma et al. (2007).} 
\end{figure}

\end{document}